\documentclass[aps,prb,superscriptaddress,twocolumn]{revtex4}
\usepackage{amsmath}
\usepackage{amsfonts}
\usepackage{graphicx}
\usepackage{color}
\usepackage{dsfont}
\usepackage{verbatim}
\usepackage[normalem]{ulem}

\begin{document}

\newcommand{\beq}{\begin{equation}}
\newcommand{\eeq}{\end{equation}}
\newcommand{\beqa}{\begin{eqnarray}}
\newcommand{\eeqa}{\end{eqnarray}}
\newcommand{\ben}{\begin{enumerate}}
\newcommand{\een}{\end{enumerate}}
\newcommand{\hs}{\hspace{1.5mm}}
\newcommand{\vs}{\vspace{0.5cm}}
\newcommand{\note}[1]{{\color{red}  #1}}
\newcommand{\notea}[1]{{\bf #1}}
\newcommand{\new}[1]{{#1}}
\newcommand{\ket}[1]{|#1 \rangle}
\newcommand{\bra}[1]{\langle #1|}

\newcommand{\tg}[1]{\textcolor{blue}{#1}}

\title{Kinetic Magnetism at the Interface Between Mott and Band Insulators}

\author{Jason Iaconis}
\affiliation{Department of Physics, University of California, Santa Barbara, CA, 93106-9530}
\author{Hiroaki Ishizuka}
\affiliation{Kavli Institute for Theoretical Physics, University of California, Santa Barbara, CA 93106-4030, USA}
\author{D. N. Sheng}
\affiliation{Department of Physics and Astronomy, California State University, Northridge, California 91330, USA}
\author{Leon Balents}
\affiliation{Kavli Institute for Theoretical Physics, University of California, Santa Barbara, CA 93106-4030, USA}

\date{\today}

\begin{abstract}
We show that the interplay of a high density two-dimensional electron gas and localized electrons in a neighboring Mott insulator leads to kinetic magnetism unique to the Mott/band insulator interface.  Our study is based upon a bilayer Hubbard model at $U=\infty$ with a potential difference between the two layers. We combine analytic results with DMRG simulations to show that magnetism, and especially ferromagnetism, is greatly enhanced relative by the proximity of the two subsystems. The results are potentially relevant to recent experiments suggesting magnetism in $R$TiO$_3$/SrTiO$_3$ heterostructures.
\end{abstract}

\maketitle
\section{Introduction}
Kinetic magnetism is a very old and elegant idea, whereby magnetic order appears solely due to the {\em motion} of the correlated itinerant electrons. The concept dates back to an argument by Nagaoka from 1966 in which he proved that ferromagnetism must exist in the Hubbard model \cite{PhysRev.147.392}.  While there have been attempts to extend these results to a wide range of models \cite{PhysRevB.80.224504, PhysRevLett.100.037202},  it has  become apparent that Nagaoka's ferromagnetism is a subtle effect which seems to be destroyed for any straightforward extension to realistic parameters \cite{PhysRevLett.69.2288}. It remains an outstanding goal to achieve this effect in an experimentally realizable model. 

In this paper, we consider the relevance of this venerable idea to artificial heterostructures of perovskite transition metal oxides.  These systems have emerged as a novel venue to explore correlated electron physics in a highly controlled environment  \cite{MRS:9125038}.  The dominant motif is that of a cubic lattice of Ti $d$ orbitals, with from 0 to one electron per site.  This is a canonical Mott material, with small overlap-induced hopping amongst neighboring $d$ orbitals, and large on-site Hubbard repulsion $U$.  Most of the physics explored experimentally originates from the so-called ``polar discontinuity''. This produces a high density two-dimensional electron gas (2DEG) at the interface between two such materials with different stacking of polar/non-polar atomic layers, ideally consisting of half an electron per planar Ti  unit cell for the case of a unit polar discontinuity.  Correlation effects may be observed for these electrons.

Such a 2DEG is in principle induced for any such polar structure, independent of other details of the constituent materials. 
 For example, it should occur at the junction between two band insulators, LaAlO$_3$/SrTiO$_3$ (LAO/STO), which is the most studied such oxide interface \cite{nature.427.423,PhysRevLett.98.216803,PhysRevLett.98.196802}.  In practice, the electron concentration observed in LAO/STO is greatly reduced from the expected value, for reasons which are not clear.  A polar discontinuity 2DEG is also expected for the interfaces between Mott insulating titanates $R$TiO$_3$ (where $R$ is a rare earth) and SrTiO$_3$ (STO), where the proper electron density has been measured experimentally\cite{1.3669402,1.4704363,1.4795612}.   These latter studies have been interpreted by treating the STO as a quantum well, viewing the $R$TiO$_3$ ($R$TO) as entirely inert and serving only to confine the electrons of the 2DEG.  When the 2DEG is sufficiently narrowly confined on both sides by $R$TO, indications of magnetism in the 2DEG are found \cite{PhysRevX.2.021014,PhysRevB.88.180403}.   In this paper, we tentatively connect this observation to the storied problem of kinetic magnetism.

A cautionary note is in order.  Ferromagnetism is ubiquitous in theoretical treatments of correlated electron materials \cite{PhysRevB.87.161119,nature.428.630}.  Most theoretical descriptions of magnetism rest on a mean field analysis, which notoriously overestimates the tendency to ferromagnetism.  The vast majority of theoretical treatments of oxide heterostructures fit into this category, including all first principles calculations of magnetism, and even sophisticated variants like dynamical {\em mean field} theory.  While such calculations are useful and suggestive, a controlled approach is desirable.

We take a distinct view of polar Mott Insulator/Band Insulator (MI/BI) interfaces.  Unlike a band insulator like LAO, the insulating $R$TO contains a very high density of correlated {\em localized} electrons, even higher than in the 2DEG.  We suggest that the mobile electrons in STO can have a dramatic effect on these localized electrons, driving magnetism.   We introduce a model which takes into account both the Mott insulating and itinerant electron degrees of freedom.  We then present a controlled limit whereby kinetic magnetism in the interface emerges independent of the bulk physics of either material.  We will further support this analysis with unbiased numerical evidence, which constitutes some of the first exact numerical results on these systems.

\section{The Model}
\begin{figure}[t]
\includegraphics[scale=0.395]{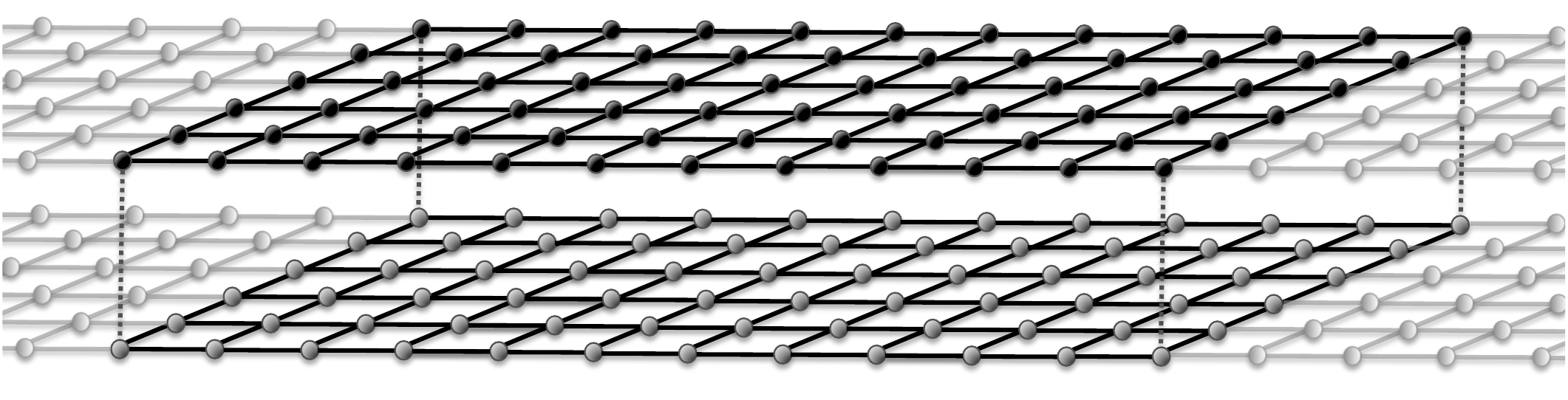}
\caption{The bilayer lattice geometry. For the numerical simulations, an elongated geometry which is optimal for the DMRG algorithm was used.  }\label{fig:fig0}
\end{figure}

We consider a minimal model that captures the physics of the MI/BI interface.  It consists of a two layer square lattice, as shown in Fig.~\ref{fig:fig0} with one layer each for the MI and BI.  If the two were decoupled, the MI would have ``exactly'' one electron per site, and the BI a lower concentration $n$ per site, where we expect $n\leq 1/2$, the maximum achievable if all the electrons in the 2DEG are in the first layer of the BI.   In reality, inter-layer hopping allows the charge to redistribute, and we include a (large) potential offset $\Delta$ to favor more electrons in the MI layer, and fix the total electron concentration to $1+n$ per two Ti sites. 
We further stress our use of an \emph{effective} single band model, which captures the effects of orbital splitting at the interface \cite{PhysRevLett.106.136803,PhysRevLett.106.166807} and includes only the electrons which make up the large majority Fermi surface \cite{apl1.4758989,PhysRevB.86.125121,cite-key}.

We model interactions by the extreme limit $U=\infty$, which forbids double occupancy.  The justification is that exchange in the $R$TO titanates is quite weak, for example the most studied materials with $R$=Sm, Gd show antiferromagnetism and ferromagnetism, respectively, with $T_c \approx 30K$ in both cases \cite{0953-8984-17-46-023}, indicating exchange $|J|$ is of order 1 meV, while $t \sim 0.3$ eV and $U \sim 4-8$ eV.  Since $J \sim t^2/U \ll t, U$, the very small exchange supports the large $U$ limit.  

With this motivation, the $U=\infty$ limit maps the Hubbard model to the so called `t-J model' with $J=0$:
\beqa
H &=& -t \sum_{\langle i j \rangle z \sigma} \mathcal{P} c_{i \sigma z}^\dagger c^{\vphantom\dagger}_{j \sigma z} \mathcal{P} - t \sum_{i \sigma} \mathcal{P}(c_{i \sigma 1}^\dagger c^{\vphantom\dagger}_{i \sigma 2} + \text{h.c.}) \mathcal{P}  \nonumber \\
&& \hspace{30mm} + \sum_{i z \sigma}(\Delta \delta_{z,1} - \mu) n_{i \sigma z}  ,\label{eqn:hubbard}
\eeqa
where $\mathcal{P} = \prod_i (1 - n_{i \uparrow} n_{i \downarrow})$.  
The only free parameters are the filling $1+n=\frac{1}{L_xL_y}\sum_{iz\sigma} \langle n_{i\sigma z}\rangle$ (or chemical potential $\mu$) and the ratio of hopping to the potential difference $(t/\Delta)$.  

The single layer, single band, $U=\infty$ Hubbard model has been the subject of many studies.  
At half filling, the system is a Mott insulator since the projection operator prevents electron hopping.  
Nagaoka famously showed in \cite{PhysRev.147.392} that when the half-filled system is doped with a  single hole, the exact ground state is the fully polarized state with maximum $S_{\text{total}}$.  
This magnetism is the result of delicate quantum effects arising from the kinetic motion of the single hole through the lattice.
The question of whether this ferromagnetism can be extended to finite doping has been attacked via mean field calculations \cite{PhysRevLett.57.1362}, variational studies \cite{PhysRevLett.86.3396,PhysRevB.41.2375}, and unbiased numerical approaches including quantum Monte Carlo \cite{PhysRevB.58.R10100} and most recently DMRG calculations \cite{PhysRevLett.108.126406}. 
While it appears that a ferromagnetic metal is stable over a finite range of filling $n$, it is clear that at lower densities ($0\le n\le 0.75$), the ground state is a paramagnetic metal.  
In this letter we will show that the bilayer model with finite band separation, $\Delta$, contains much richer magnetic structure at all filling densities.
In particular, at large band separation we are able to stabilize Nagaoka's ferromagnetism over a wide range of electron densities $n$.

\section{  Perturbative Regime ($\Delta \gg t$) }
In the limit of large $\Delta$ we can demonstrate analytic control over the model. 
At $\Delta = \infty$, the two layers are completely decoupled, where the upper layer is a degenerate spin system and the bottom layer behaves according to the results of Ref.\cite{PhysRevLett.108.126406}.
 In particular, for $\langle n \rangle < \frac{3}{4}$,  the bottom layer is a paramagnetic metal. 
  If we now tune away from $\Delta = \infty$, we can derive an effective low-energy Hamiltonian perturbatively in $(t/\Delta)$.  To lowest order in the perturbative expansion,
\beqa
\mathcal{H}_{eff} &=& H_{00} + H_{01}\frac{1}{E-H_{11}}H_{10} .
\eeqa
where $H_{10}$ hops an electron from the top to the bottom layer, and $H_{01}$ brings us back into the ground state subspace of no holes in the top layer.
Assume that the density in the bottom layer is such that there is a paramagnetic metal. 
In this case, the virtual contribution to the energy when there is a single hole in the top layer, by Nagaoka's result, is minimized when the top layer is a fully polarized ferromagnet.  
Then for nearly all densities at large $\Delta$, the degenerate groundstate subspace splits in a way that causes the ferromagnetic state to become the true ground state.
However, this argument breaks down at the lowest electron densities, since here there are no electrons present at different spatial sites to fill the virtual hole in the top layer.  The electron is then effectively localized and the ferromagnetism is lost.

We will now make this argument precise.
We expand the Hamiltonian to order $(t/\Delta)^3$, by using the identity
\beqa
\frac{1}{\omega  - H} &=& \frac{1}{\omega} + \frac{1}{\omega}H \frac{1}{\omega-H}. 
\eeqa
 The lowest order effect, which occurs at order $(t/\Delta)^2$, is
\beqa
\hspace{-6mm} H^{\prime(1)} &=& - \frac{t^3}{\Delta^2} \sum_{ \langle i j \rangle} \sum_{ \sigma \sigma^\prime \sigma^{\prime \prime}} c_{i2\sigma}^\dagger c^{\vphantom\dagger}_{i1\sigma}c_{j2\sigma^\prime}^\dagger c^{\vphantom\dagger}_{i 2\sigma^\prime} c_{j 1 \sigma^{\prime \prime}}^\dagger c^{\vphantom\dagger}_{j 2 \sigma^{\prime \prime}} \nonumber \\
 &&\hspace{-4mm} = -\frac{t^3}{\Delta^2}  \sum_{\langle i j \rangle}  \sum_{\alpha \beta} \left [ \vec{S}_i \cdot \vec{S}_j  \, \delta_{\alpha \beta} + \frac{1}{2}(\vec{S}_i + \vec{S}_j)\cdot \vec{\sigma}_{\alpha \beta} \right . \nonumber\\
&& \hspace{14mm} \left . \phantom{\frac{1}{2}}- \, \, i (\vec{S}_i \times \vec{S}_j)\cdot \vec{\sigma}_{\alpha \beta} \right ] \mathcal{P} c_{j1\alpha}^\dagger c^{\vphantom\dagger}_{i 1 \beta} \mathcal{P}  \label{Hprime1}
\eeqa
This expression suggests an obvious way to decouple the terms at the mean field level, by taking expectation values of operators in the same layer.  This leaves us with an effective spin model for the upper layer and a doped electron system in the bottom layer.  The antisymmetric form of the third term in Eq.~\eqref{Hprime1} implies we can ignore its mean field effect at this order in perturbation theory. The first term then gives the effective interaction in the upper layer as a ferromagnetic Heisenberg interaction with $J_{FM} = -t^3 \langle c_{j}^\dagger c^{\vphantom\dagger}_i \rangle / \Delta^2 \sim (t^3 n) / \Delta^2$.
\begin{figure}[t]
\includegraphics[scale=0.60]{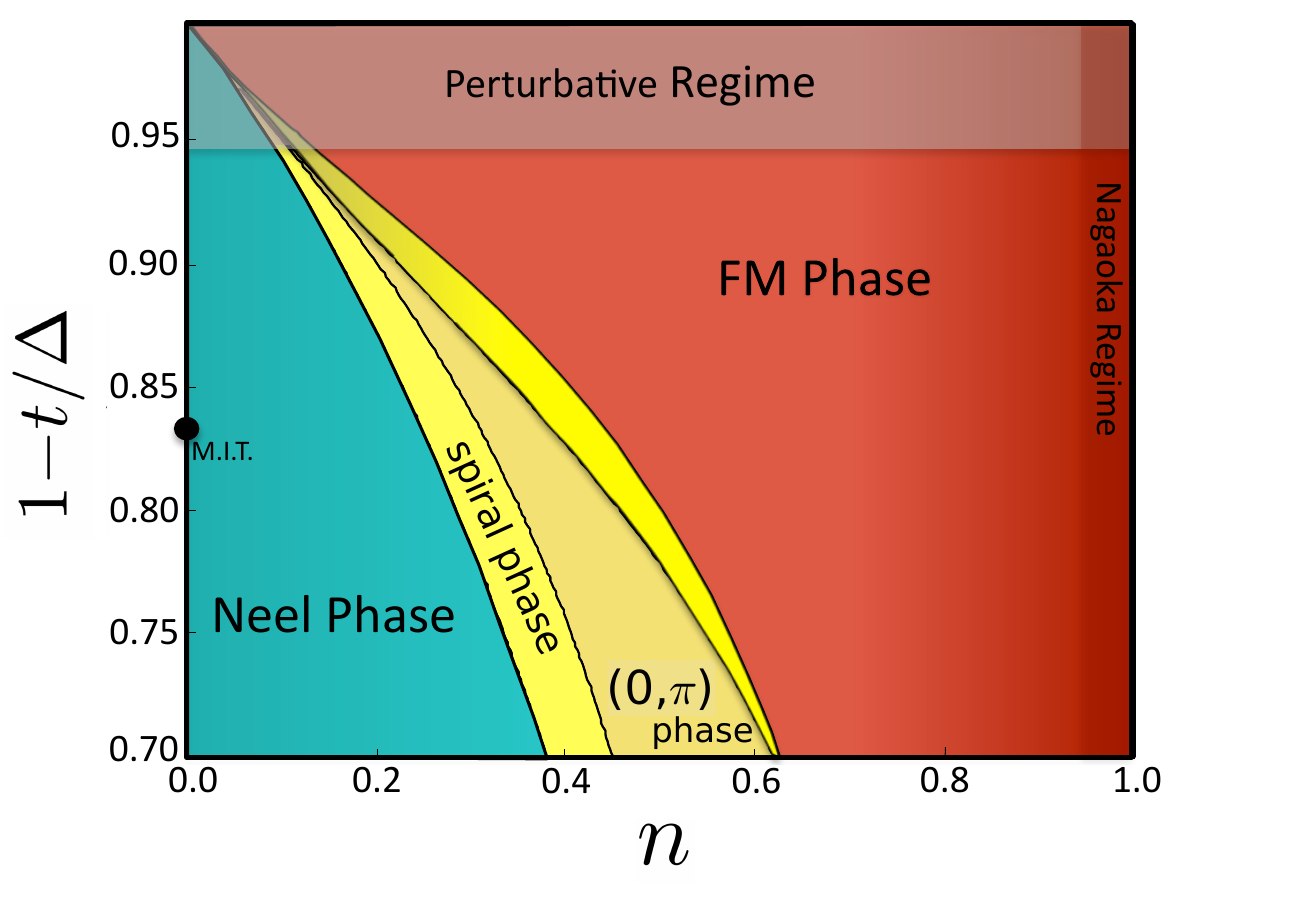}
\caption{We embed the classical phases of the $J_1-J_2-J_3$ Heisenberg model into our bilayer Hubbard phase diagram, using the form of $J_1, J_2$ and $J_3$ given in the text. These results become rigorous for large $\Delta/t$.  The highest densities are ferromagnetic by Nagaoka's theorem.}\label{fig:fig1}
\end{figure}

At zero temperature, the energy can be lowered at the mean-field level if the upper layer forms a fully polarized ferromagnet.
The second term of Eq.~\eqref{Hprime1} then provides an effective magnetic field in the ordering direction of the upper layer spins. If $n \rightarrow 0$ then $J_{FM} \rightarrow 0$ also, and we must look at the next order in perturbation theory.
 At this order, we derive additional antiferromagnetic interactions which compete with the lowest order term.  These can be written as
\beqa
H^{\prime (2)} &=& \frac{4 t^4}{\Delta^3} \langle (1-n) \rangle^2 \sum_{\langle i j \rangle} \vec{S}_i \cdot \vec{S}_j \\
&& \hspace{-5mm} + \frac{t^4}{\Delta^3} \sum_{\langle \langle i j k \rangle \rangle} \langle c_{i}^\dagger c^{\vphantom\dagger}_k \rangle \left[ (\vec{S}_i+\vec{S}_j + \vec{S}_k)\cdot (\vec{S}_i+\vec{S}_j + \vec{S}_k )\right] \nonumber
\eeqa
where $\langle \langle i j k \rangle \rangle$ implies the sum is over all connected clusters of 3 sites on the same layer. This  therefore describes next and third nearest-neighbor interactions.

When $(t/\Delta)$ is small, we can treat the upper layer of our bilayer model as a spin system with nearest, next-nearest and third-nearest neighbor interactions.
The resulting effective Hamiltonian is equivalent to the so called $J_1-J_2-J_3$ Heisenberg model.
 The parameters, $J_1$, $J_2$, and $J_3$, are related to the original parameters $t$ and $\Delta$ via the results of the previous section.  $J_1$ can thus be either ferromagnetic (FM) or antiferromagnetic (AFM), but $J_2$ and $J_3$ are always antiferromagnetic. 
Away from $n \approx 1$ and $\Delta/t \approx \infty$, this effective Hamiltonian is frustrated.  
While a full quantum solution for such a model on the square lattice is still lacking, the classical solution is well understood \cite{PhysRevLett.66.1773, PhysRevB.47.8769, PhysRevB.44.392, PhysRevB.42.6283,PhysRevB.40.10801}. 
We embed this classical solution in the $t-n$ phase diagram in Fig.~\ref{fig:fig1}. 
There are four distinct phases. 
When $\Delta$ is large, $J_1$ is large and positive and the ground state is a simple ferromagnet.
At lower densities, $J_1$ is large and negative and the system is in a N\'eel phase.
Between these limits, the two contributions to $J_1$ nearly cancel, and the second and third neighbor terms become important.
In these cases the ground state is either a striped phase with wave-vector peaked at $(0,\pi)$, or a spiral phase which interpolates between the striped and FM or the striped and N\'eel phases.  
We note that, quantum mechanically, the regime of competing exchanges might host another exotic state such as a valence bond solid or quantum spin liquid.

\section{ Instability of Ferromagnetism }
\begin{figure}[t]
\includegraphics[scale=0.50]{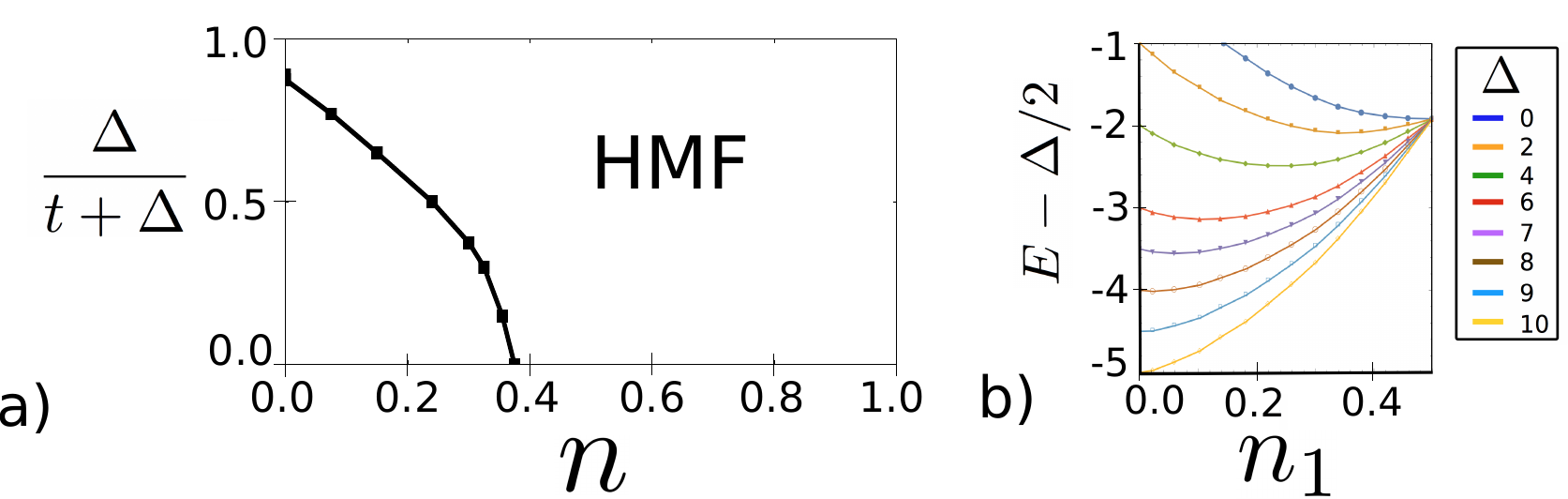}
\caption{Results of (a) the variational calculation and (b) the Gutzwiller approximation. (a) The stability of the fully polarized FM state to the Gutzwiller projected trial state with a single flipped spin. The FM state becomes unstable inside the area bounded by the solid line. (b) Ground state energy with respect to $n_1 = \sum_\sigma\langle n_{i\sigma 1}\rangle$ calculated by Gutzwiller approximation at $n=0$.}
\label{fig:fig3}
\end{figure}
We next study the instability of ferromagnetism using a variational method. 
Since double occupancy is forbidden automatically in the fully polarized or `half-metallic ferromagnet' (HMF) state due to fermi statistics, its energy can be calculated exactly.  We then can prove that this state is {\em not} the ground state if we find any state with lower variational energy. We consider the same trial state as in Ref.~\cite{PhysRevB.41.2375},
\beqa
\ket{\psi} &=& \mathcal{P} \psi_\downarrow^\dagger \ket{FM^\prime} \\
\psi_{\downarrow} &=& \sum_{i \alpha} \xi_{i \alpha} c_{i \alpha \downarrow}^\dagger
\eeqa
where $\ket{FM^\prime}= c_{\vec{k}_F}\ket{FM}$ is the fully polarized metal with one less electron than $\ket{\psi}$, and $\mathcal{P}$ is the Gutzwiller projection operator which forbids double occupancy of any site, and $\xi_{i \alpha}$ are variational parameters.  

Further details of the variational calculation are given in the appendixes. The results are shown in Fig.~\ref{fig:fig3}(a). The trend is towards increased ferromagnetism for larger $\Delta$, in agreement with the perturbative results. This implies that for large enough hole concentrations, the Nagaoka state is unstable to flipping an electron spin, consistent with the intuitive picture.
 The instability, however, weakens for larger $\Delta$ and we could not find an unstable region for $\Delta\gtrsim6.5$.

We next turn our attention to the metal-insulator transition (MIT) at $n=0$ with increasing $\Delta$. To investigate the MIT, we here study the model in Eq.~\ref{eqn:hubbard} by the Gutzwiller approximation assuming a paramagnetic solution \cite{PhysRev.137.A1726,PhysRevLett.10.159,Ogawa01021975,RevModPhys.56.99}. In this framework, the MIT is characterized by the absence of electrons in the bottom layer. As is shown in Fig.~\ref{fig:fig3}(b), this occurs at $\Delta \simeq 8t$. This is consistent with our DMRG results, where we find the single particle excitation gap $E_g = E(n+1) - 2E(n) + E(n-1)$ becomes nonzero continuously in the 4 leg ladder at $\Delta = 6t$.

\section{Numerical Results }
\begin{figure}[t]
\includegraphics[scale=0.370]{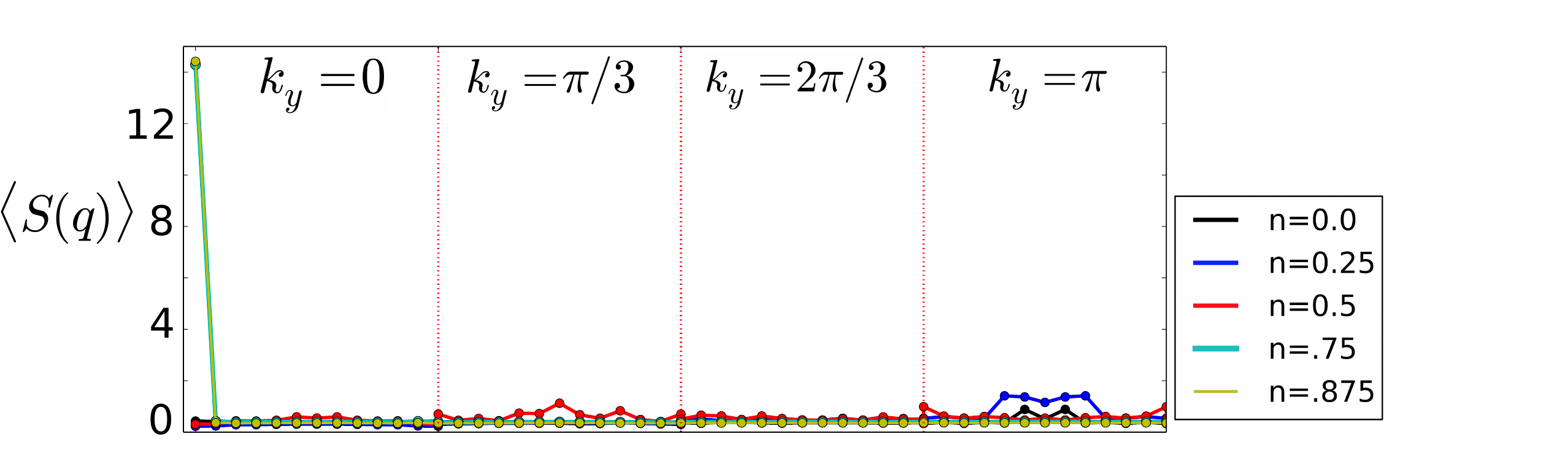} \\
\includegraphics[scale=0.370]{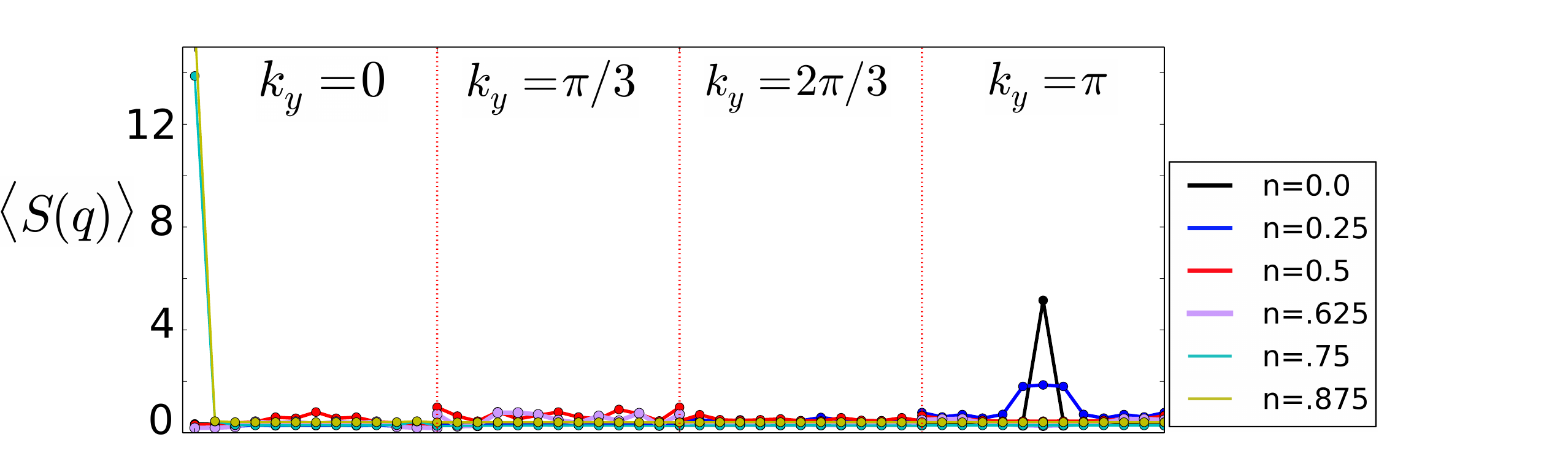}\\
\includegraphics[scale=0.370]{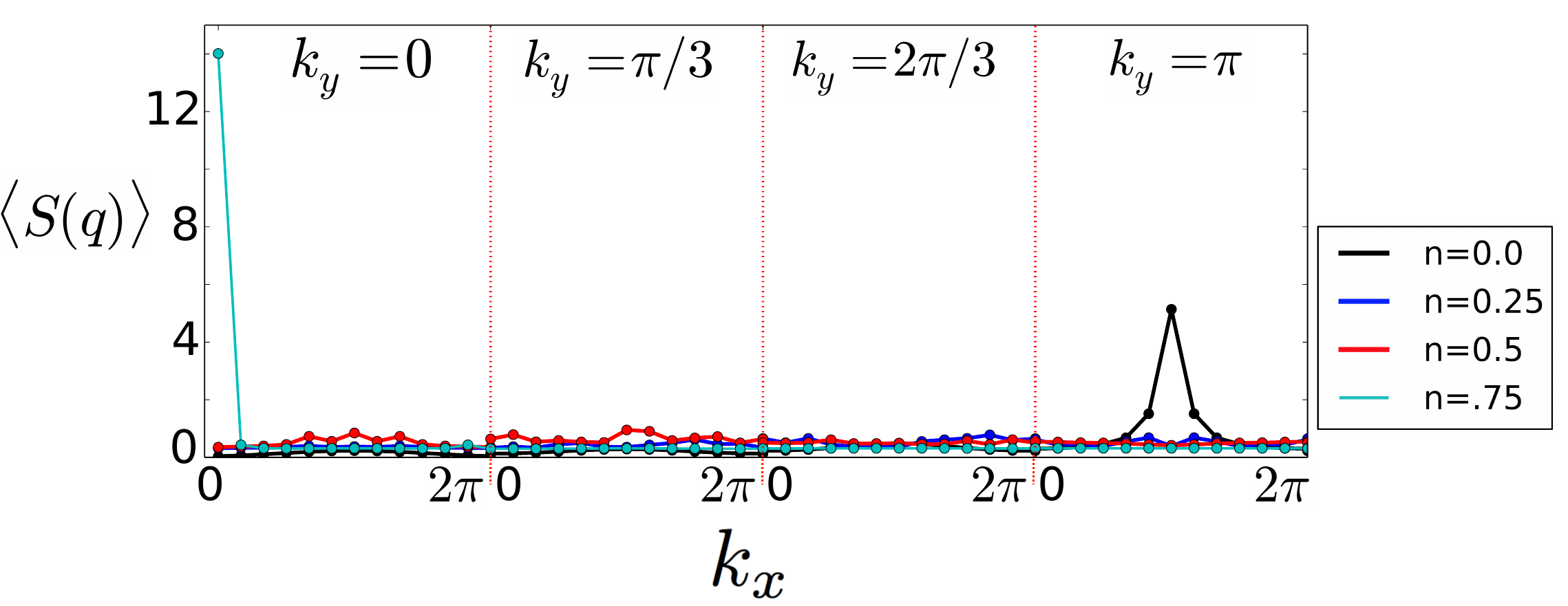}
\caption{The structure factors for the 6x24x2 system for ({\it from top to bottom}) $\Delta = 4,6$ and $10$.  The highest densities are always ferromagnetic. N\'eel order becomes more stable for smaller $\Delta$, and the intermediate regions show no strong peaks.}
\label{fig:fig2}
\end{figure}
We will now demonstrate the consistency of our analytic arguments with unbiased numerical results.  
We performed a series of DMRG calculations on bilayer systems of up to six leg ladders. 
The total number of sites is then 2$\times$24$\times$6.  We keep 4000 to 6000 states and the truncation error is
of the order of $10^{-6}$ in the ferromagnetic phase,  but increases to $10^{-4}$ in the paramagnetic phase.

  In our DMRG set up, we first combine the two layer system into an effective
one layer system.  The new rung index is $R_x^{new}=2*(R_x-1)+\tau$ where
$R_x$ is the rung index of each layer, and $\tau=1, 2$ is the layer index.
The DMRG study for the effective one layer system follows the standard DMRG
for a cylinder. The two layer system is reflected in the Hamiltonian of the 
effective one layer system (which has a doubled unit cell along $x$, besides
the open boundary conditions we used).   The convergence  crucially depends on which state
we obtain in the different parameter regimes.  For the ferromagnetic ground state,
we are  able to go to a large total $S_z$ subspace,  which has a substantially reduced Hilbert space dimension.
For other phases (the metallic phase in particular),  DMRG indeed has a large truncation error and the results are 
not converged for such a bilayer system (which is not the focus of our study). 

Due to the difficulty of the simulations, we limit our search over phase space to values of $\Delta/t = 4,6$ and $10$ and the fillings $n = 0, 0.25,0.5,0.75$ and $0.875$.  We focus mainly on the spin-spin structure factor $S(q) = \sum_{j}e^{i \vec{q}\cdot \vec{x}_{ij}} \langle \vec{S}_i \cdot \vec{S}_j \rangle$.  These results are summarized in Fig.~\ref{fig:fig2}.

For $n \ge 0.75$ and all $\Delta \ge 4$, we find very large peaks in the structure factor at wave-vector $(q_x,q_y) = (0,0)$, consistent with a  nearly fully polarized ground state.  In all cases, the total spin $S$ satisfies $S \ge 0.90 S_{\text{max}}$.  In fact, for $\{n=0.875;\Delta = 4,6\}$, we find $S \ge 0.98 S_{\text{max}}$. Note that this does extend the range of ferromagnetism from the results of Ref.~\cite{PhysRevLett.108.126406}, which find the HMF in the single layer model only up to fillings $n = 0.8$.  

At the lowest densities $n=0$ and $n=0.25$, we find very strong agreement with our predicted results from perturbation theory.  At $\{n = 0,\Delta = 6,10\}$, there are large peaks in the structure factor at the  $(\pi, \pi)$ wave-vector.  This suggests the presence of strong staggered magnetism consistent with a N\'eel phase.  For $n=0.25$, we find a smaller N\'eel peak at $\Delta = 6$, which then disappears as $\Delta$ is increased to $\Delta=10$.  This is again consistent with our perturbative results which suggest that AFM exchange is stronger for smaller $\Delta$.  

From the classical phase diagram of the effective perturbative spin model we expect striped or spiral order to interpolate between the N\'eel and FM phases.  Our results on 6-leg ladders for $\{n=0.5; \Delta = 6,10\}$ and $\{n=0.25; \Delta =10\}$ show no strong evidence of magnetic order.  We do observe small peaks which may presage spiral or stripe order in larger systems.  

We provide further evidence for this magnetic ordering in the appendix, by calculating the momentum distribution function.

Finally, we would like to stress that although ferromagnetism occurs over a smaller range of densities in the numerical results, our perturbative phase diagram must be exactly correct for sufficiently large $\Delta$.
 However, it is possible that the range of $\Delta$ accessible in our simulations is not large enough to see the full extent of this effect. 

\section{ Conclusions } In closing we note that a more faithful representation of the oxide interface would include additional complications such as multiple $t_{2g}$ electron orbitals and super-exchange interaction $J$.  For $R$=Gd,Sm, which are strongly distorted from the cubic structure, the intrinsic $J$ is so weak that the kinetic mechanism described here is dominant or at least competitive with $J$, and orbital splittings are large. In general, however, these effects may work to stabilize certain types of magnetic order \cite{natphys9.10}.
 For example, directional hopping of the $t_{2g}$ orbitals may act to favor ferromagnetism for smaller values of $\Delta$.
 Our model avoids these complications by considering a simple limit where only the filling $n$ and band offset $\Delta$ are free parameters, yet nevertheless provides a picture of the physics.  We suggest that scattering experiments to directly probe the magnetic order in the vicinity of these interfaces would be the most direct test of our theoretical predictions.

\begin{acknowledgments} 
We gratefully acknowledge discussions and inspiration from Susanne Stemmer and Stephen Wilson.
This work was supported by the Army Research Office (J.I. and L.B.), Grant No. W911-NF-14-1-0379 and NSERC of Canada (J.I.). This work was partially supported by the MRSEC Program of the National Science Foundation under Award No. DMR 1121053 (H.I.) and by the JSPS Postdoctoral Fellowships for Research Abroad (H.I.).  D.N.S. was supported by the NSF, grant DMR-1408560.
\end{acknowledgments}

\appendix 
\section{Momentum Distribution Function}

In this section, we present the momentum distribution function as calculated using DMRG for a 6-leg ladder and $\Delta = 4$.  We can estimate the position of the Fermi surface from the apparent discontinuity in the distribution function.  We only show the results for $\Delta = 4$.  We again exclude $2L_yL_x/4$ sites on each end of the ladder for the purpose of reducing boundary effects.

\begin{figure}[h]
\includegraphics[scale = 0.570]{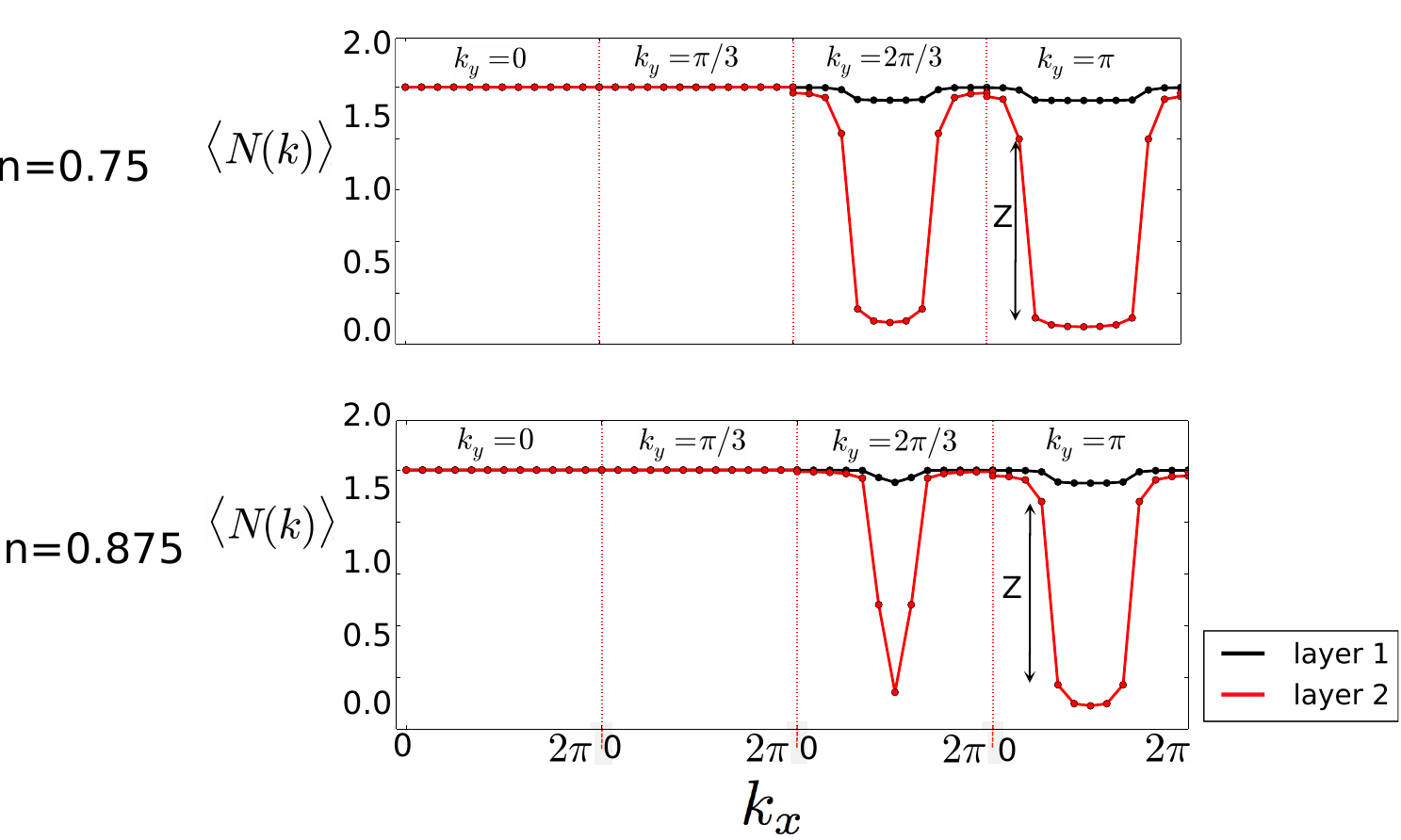}
\caption{The momentum distribution function of the 6-leg bilayer model, for $\Delta = 4$ and  high electron densities.  The Luttinger volume is consistent with a polarized state.  We show only the $k_y$ cuts that are not related by inversion symmetry.}\label{fig:fignk1}
\end{figure}

For the largest two densities, $n=0.75$ and $n=0.875$, as seen in Fig.~\ref{fig:fignk1}, the volume enclosed by the Fermi surface is $\text{Vol}/(2\pi)^2 = 0.75$ and $0.875$ respectively.  This Luttinger volume is consistent with a polarized Fermi gas, whereby the upper band is completely filled and every electron fills a different momentum state in the lower band.  The discontinuity gives the quasiparticle residue.  We see that this value is slightly less than that of a noninteracting polarized Fermi gas, signaling the fact that the ground state here is nearly fully polarized with a few flipped spins (i.e. $S > 0.90 S_{max}$).

\begin{figure}[t]
\includegraphics[scale = 0.575]{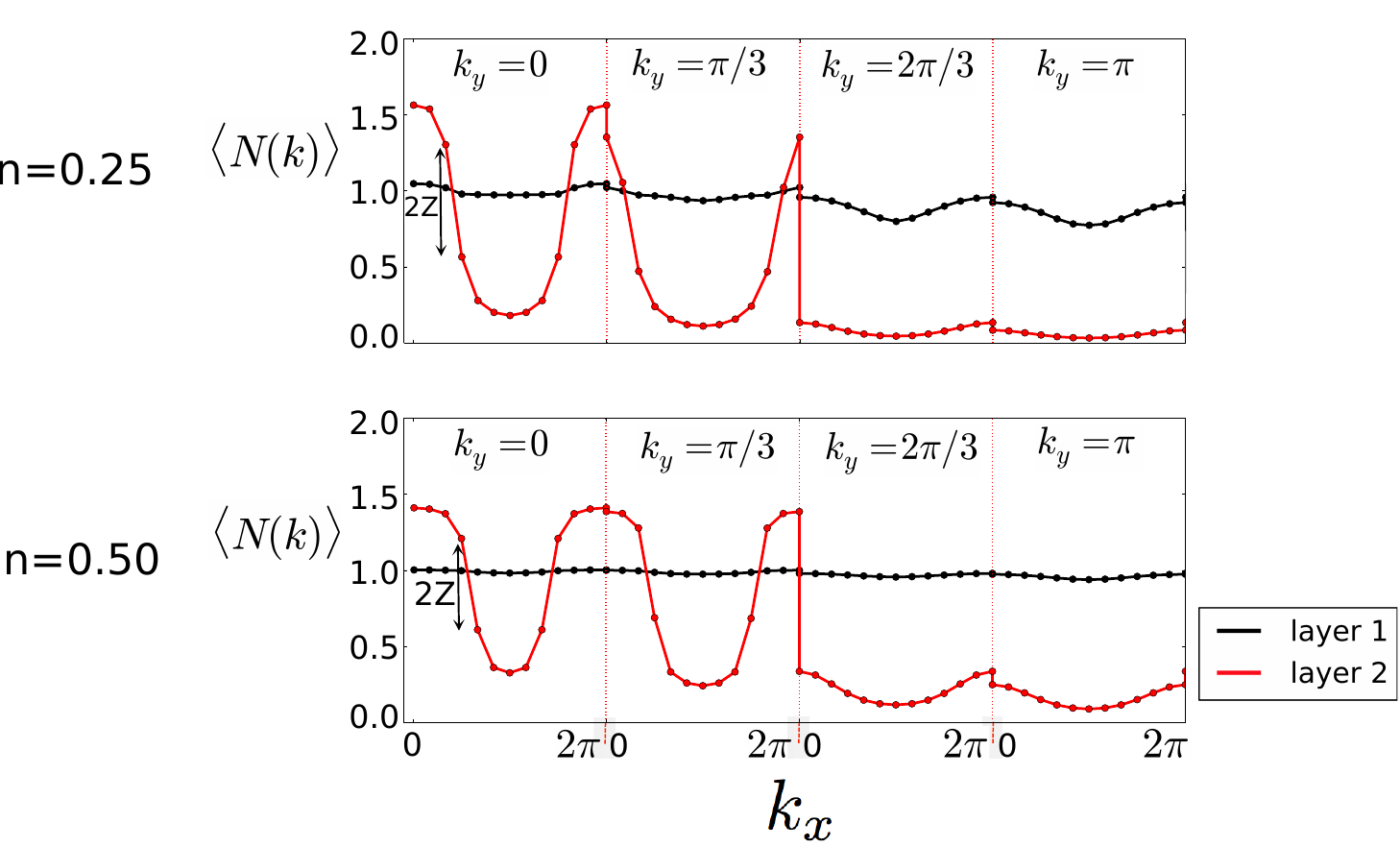}
\caption{The momentum distribution function for $n=0.25$ and $n = 0.5$.  With the smaller Luttinger volume, the discontinuity now gives twice the quasiparticle residue. }
\label{fig:nk2}
\end{figure}

In Fig.~\ref{fig:nk2} shows the same calculation for $n = 0.25$ and $n=0.5$.   The top layer is filled very uniformly, with all momentum states occupied. The Fermi surface then encloses a volume equal to half that of the number of electrons in the bottom layer.  
 This is consistent with the small Fermi volume of an unpolarized Fermi liquid. 
 For $n=0.25$ the structure factor indicates  the presence of N\'eel order, which implies there is a doubling of the unit cell. This allows the upper layer electrons to form a completely filled band so that only the lower layer electrons contribute to the Luttinger volume.
  For $n=0.5$, we find the same Luttinger volume as the $n=0.25$ case. Here, however, the structure factor showed no evidence of magnetic order.  The fact that only the lower layer electrons contribute to the Fermi volume, however, rules out the possibility of a trivial paramagnetic metal.  If the absence of magnetic order survived to the thermodynamic limit, this would be the $FL^*$ phase, which describes a quantum paramagnetic metal with a `small' Fermi volume.
We also see that the quasiparticle residue is much smaller in this regime, indicating that the ground state here is a strongly interacting state.

\begin{figure}[h]
\includegraphics[scale = 0.57]{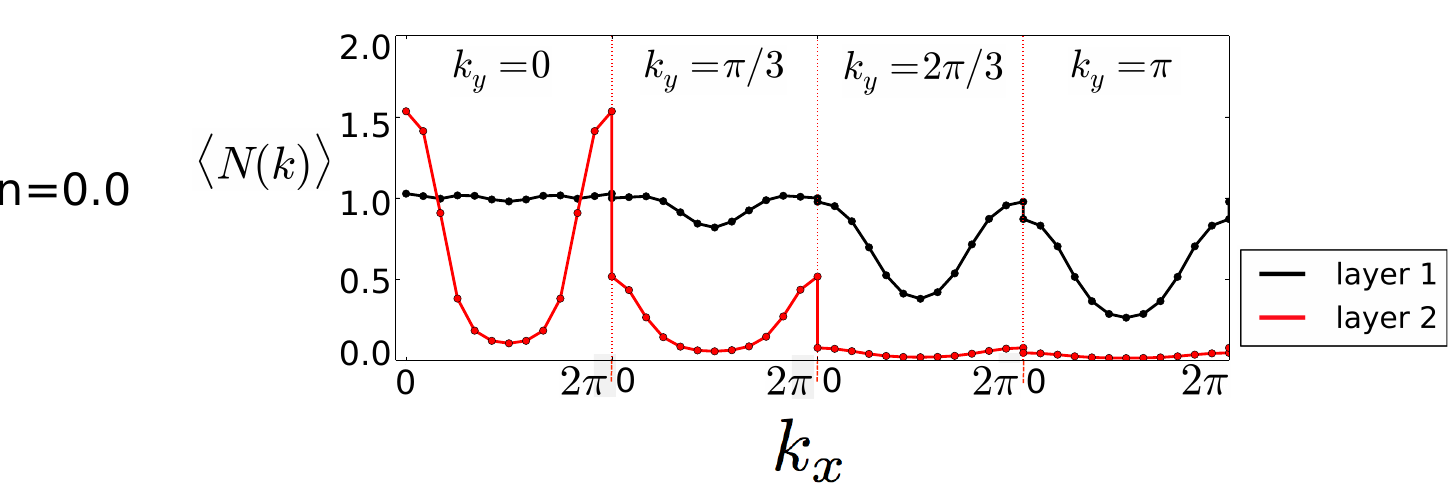}
\caption{When $n=0$, and at $\Delta= 4$, there appear to be no sharp discontinuities. }
\label{fig:nk3}
\end{figure}

Finally, we look at the case when $n = 0$.  For this filling, there exists a metal insulator transition at a critical $\Delta_c$. When $\Delta = 4$, we are on the metallic side of this transition.  From Fig.~\ref{fig:nk3}, we see no apparent discontinuities in the momentum distribution.  This could indicate the existence of a non-Fermi liquid ground state in this parameter range.  Note that although we are on the metallic side of the MIT, the structure factor shows a small ($\pi,\pi)$ peak.  This type of spin-density wave transition coupled to a Fermi surface has been studied extensively in the literature and is strongly suspected to show non-Fermi liquid behavior. 

\begin{center}
{\bf Appendix B: Variational Results}
\end{center}
\vspace{1pc}

In the main text, we show the instability of the fully-polarized ferromagnetic state by comparing the energy to a trial state.
Here, we present details of the method we used.

In the variational calculation, we consider a trial state
\beqa
\ket{\psi} &=& \mathcal{P} \psi_\downarrow^\dagger \ket{FM^\prime} \\
\psi_{\downarrow} &=& \sum_{i \alpha} \xi_{i \alpha} c_{i \alpha \downarrow}^\dagger
\eeqa
where $\ket{FM^\prime}= c_{\vec{k}_F}\ket{FM}$ is the fully polarized metal with one less electron than $\ket{\psi}$, and $\mathcal{P}$ is the Gutzwiller projection operator which forbids double occupancy of any site, and $\xi_{i \alpha}$ are variational parameters.  

With some calculation, we obtain 
\beqa
\varepsilon_\downarrow = \frac{\bra{\psi}H -E_\text{FM} \ket{\psi}}{\langle \psi | \psi \rangle}  = \sum_k \hat{\xi}_k h_k \hat{\xi}_k,
\eeqa
where $E_\text{FM}$ is the ground state energy for the fully-polarized state and $\hat{\xi}_{k}=(\xi_{k 1},\xi_{k 2})$ with $\xi_{k \alpha} = \sum_i \xi_{i \alpha} \exp(i \vec{k}\cdot \vec{r}_i)$, $\alpha=1,2$.  Additionally, $h_{\bf{k}}$ is a $2\times2$ effective Hamiltonian whose explicit form is
\beqa
h_{\bf{k}}= \left( \begin{array}{cc} -\tilde{t}_0 \epsilon_{\bf{k}} - T_0 &  -\tilde{t}^\prime\\ 
-\tilde{t}^\prime & \tilde{t}_1 \epsilon_{\bf{k}}+ \tilde{\Delta} - T_1 \end{array}  \right ),
\eeqa
with
\beqa
\tilde{t}_a &= &\frac{t}{R} \langle (1-n_{ia\uparrow})(1-n_{ja\downarrow}) \rangle \\
\tilde{t}^\prime &=& \frac{t^\prime}{R} \langle (1-n_{i1\uparrow})(1-n_{j2\downarrow}) \rangle \\
\tilde{\Delta} &=& \frac{\Delta }{R} \\
T_a &=& \frac{1}{R} \sum_{j,b} t_{ia,jb} \langle c_{i a \uparrow}^\dagger c_{j b \uparrow} \rangle
\eeqa
and where $R^2 = \bra{\psi}\psi \rangle$.
The optimal variational parameters are then just given via the solution of this single particle problem, and $\epsilon_\downarrow$ is given by the smallest eigenvalue of $h_{\bf{k}}$.

\bibliography{base}

\end{document}